# Chemical reactions in the nitrogen-acetone ice induced by cosmic ray analogs: relevance for the Solar System

A. L. F. de Barros[1,2], D. P. P. Andrade[3], E. F. da Silveira[4], K. F. Alcantara[4], P. Boduch[5], H. Rothard[5]

[1] *Departamento de Física, Centro Federal de Educação Tecnológica Celso Suckow da Fonseca, Av. Maracanã 229, 20271-110 Rio de Janeiro, RJ, Brazil*

[2] *NASA Ames Research Center, Mail Stop 245-6, Moffett Field, CA 94035-1000, USA*

[3] *Universidade Federal do Rio de Janeiro, Observatorio do Valongo, Ladeira Pedro Antonio, 43, Rio de Janeiro, Brazil*

[4] *Departamento de Física, Pontifícia Universidade Católica do Rio de Janeiro, Rua Marquês de São Vicente 225, 22451-900, Rio de Janeiro, RJ, Brazil*

[5] *Centre de Recherche sur les Ions, les Matériaux et la Photonique Normandie Univ, ENSICAEN, UNICAEN, CEA, CNRS, CIMAP, 14000 Caen, France*

## ABSTRACT

The radiolysis of a 10:1 nitrogen:acetone mixture, condensed at 11 K, by 40 MeV $^{58}$Ni$^{11+}$ ions is studied. These results are representative of studies concerning solar system objects exposed to cosmic rays. In the Kuiper Belt, region of Trans-Neptunian Objects (TNOs), acetone, $N_2$ and other small molecules were detected and may be present on icy surfaces. Bombardment by cosmic rays triggers chemical reactions leading to synthesis of larger molecules. In this work, destruction cross sections of acetone and nitrogen molecules in solid phase are determined from a sequence of infrared spectra obtained at increasing ion beam fluence. The results are analyzed and compared with those of previous experiments performed with pure acetone. It is observed that the $N_2$ column density decreases very fast, suggesting that nitrogen quickly leaves a porous sample under irradiation. The behavior of acetone in the mixture confirms that the ice formed by deposition of the vapor mixture is more porous than that of pure acetone ice. The most abundant molecular species formed from the mixture during irradiation are: $C_3H_6$, $C_2H_6$, $N_3$, CO, $CH_4$ and $CO_2$. Some N-bearing species are also formed, but with low production yield. Comparing with pure acetone results, it is seen that dissolving acetone in nitrogen affects the formation cross sections of the new species: $CH_4$, $CO_2$ and $H_2CO$, for example, have formation cross section smaller than the respective values for irradiated pure acetone, while those for CO and $C_2H_6$ species are higher. This fact can explain the presence of $C_2H_6$ even in regions on Pluto where $CH_4$ is not pure, but diluted in a $N_2$ matrix together with less abundant species. These results also show the formation of more complex molecules, such as HNCO, acetic acid and, possibility, glycine. The production of these complex molecules suggests the formation of small prebiotic species in objects beyond Neptune from acetone diluted in a $N_2$ matrix irradiated by cosmic rays. Formation cross sections for the new formed species are determined.

# 1 INTRODUCTION

Acetone ($CH_3COCH_3$) was observed by the first time in the interstellar medium by Combes et al. (1987) through the molecular cloud Sgr B2. Recently, acetone has also been detected in comets; the $N_2$ molecule is a common species in these objects, as well as in dwarf planets. Comets were observed through optical and radio spectra, from which a number of molecular species have been identified. Studies in the ultraviolet showed the presence of species such as CN, $C_2$, $C_3$ (Newburn et al. 1994; Cochran et al. 1987), which can be fragments of C and N-rich molecules. Afterward, other works showed that several observed molecular species are formed by C, O, S and N atoms (Fink 2009; Langland-Shula & Smith 2011; Cochran et al. 2012; Schleicher et al. 2014), reinforcing the idea that comets are rich in species containing these atoms.

In the infrared (IR) absorption range, molecular species such as $H_2O$, CO, $CO_2$, $CH_4$, $C_2H_2$, $H_2CO$, $NH_2$, $NH_3$, $CH_3OH$ and HCN were detected in TNOs. This list increases if species observed by radio observations, such as $CH_3CN$, $HC_3N$, HCN, HCOOH, HNCO and $H_2S$, are considered. Besides these molecules, $SO_2$, $H_2CS$, $NH_2CHO$, $CH_3CHO$, $HO-CH_2CH_2-OH$ (ethylene glycol), $CH_3COOH$ (acetic acid) and its isomer $HCOOCH_3$ (methyl formate) were also detected in the comet C/1995 O1 (Hale-Bopp), (Biver et al. 2002; Bockelee-Morvan et al. 2000; Crovisier et al. 2004).

It is believed that comets originate from two sources: Oort cloud (long-period comets) and Kuiper Belt (short-period comets). The Kuiper Belt is a disk-shaped region filled with icy bodies beyond the orbit of Neptune (around 30-60 AU from the Sun). These icy bodies in the Kuiper Belt are called Kuiper Belt Objects (KBOs for short), Edgeworth-Kuiper Belt objects or sometimes Transneptunian Objects (TNOs for short). Gravitational interactions with giant planets can disturb the orbit of a KBO and throw it into the inner solar system. Even today, with large telescopes, only the surface compositions of a few KBOs are known. The chemical composition does not represent the composition of the entire body because most of objects are small and far away. The problem of determining chemical composition of KBO objects is complex because of their sizes and distance. Laboratory simulations are a valuable alternative for the research on the chemistry of ices and minerals occurring in bodies staying in early outer solar system. This motivates the current work.

Frozen objects located at heliocentric distances larger than 30 AU must have been formed in the outer layers of the solar disk or in the ISM. The ice existing in this reservoir is amorphous and gases can be trapped in its pores (Bar-Nun et al. 2007). Dynamical models point to the direction that Jupiter Family Comets and Oort Clouds comets originated from the same parent population, located in the primordial trans-Neptunian disk (10-40 AU region).

According to Luu et al. (1994) and Cooper et al. (1998), the mantle formation on KBO seems to be processed by the anomalous cosmic rays (ACR), constituted by interstellar neutral gas which is ionized close to the sun by photo-ionization or by charge exchange with the solar wind. Ionized particles are dragged by the solar wind to the periphery of the heliosphere, where turbulent magnetic fields may throw the particles back inside the heliosphere (Pesses et al. 1981; Izzo et al. 2014). The Voyager 2 spacecraft obtained the energy spectra of C, N and O at 23 AU during the solar minimum in 1987 and showed that the maximum flux of these species is around 3-6 MeV/nucleon. He ions have a maximum around 20 MeV/nucleon. At energies of about 50 MeV/nucleon, galactic cosmic ray dominate.

Recently, the Cometary Sampling and Composition (COSAC), a device aboard of Rosetta's Philae lander, analyzed in situ organic molecules on the surface of the comet 67P/Churyumov-Gerasimenko, a Jupiter-family comet, originally from the Kuiper Belt. The COSAC consists of a time-of-flight spectrometer (TOF-MS) and a gas chromatograph (GC-MS). The TOF-MS identified diverse organic compounds, among them several nitrogen-bearing species, including four new compounds that had not been reported before in comets: methyl isocyanate ($CH_3NCO$), acetone, propionaldehyde ($CH_3CH_2CHO$, an acetone isomer), and acetamide ($C_3CONH_2$) (Goesmann et al. 2015). Although the presence of $N_2$ has not been confirmed in the TOF-MS spectra, the appearance of complex species containing $N_2$, such as HCN, $CH_3CN$, HNCO, $HCONH_2$, $CH_3NH_2$, $C_2H_5NH_2$ (ethylamine), $CH_3NCO$ and $CH_3CONH_2$ (acetamide) suggests that nitrogen was present since the beginning of the cometary formation. Because the $N_2$ sublimation temperature is very low (T ~ 23 K at $10^{-8}$ mbar pressure), it is expected that this species has already been sublimated completely and only the products of reactions involving this precursor remain.

In this way, studies aiming the identification of products generated by chemical reactions between acetone and $N_2$ are relevant, which is one of the main objectives of the current work. Some of these species are particularly important for the origin of life; HCN, for example, has a crucial role in the synthesis of amino acids and nucleobases, whereas $HCONH_2$ and $CH_3CONH_2$ are important in the phosphorylation of nucleoside into nucleotides (Goesmann et al. 2015).

The radiolysis induced by energetic ion bombardment of the $N_2:CH_3COCH_3$ ice mixture (10:1) is here studied by Fourier Transform InfraRed (FTIR) spectroscopy. The destruction cross sections of the precursor molecules in the mixture after irradiation with 40 MeV $^{58}Ni^{11+}$ are determined, as well as the formation and destruction cross sections of the daughter molecular species.

## 2 EXPERIMENTAL SETUP

The sample was irradiated with 40 MeV $^{58}Ni^{11+}$ (E ~ 0.70 MeV.u$^{-1}$) heavy ions provided by the Grand Accélérateur National d'Ions Lourds (GANIL), in Caen, France. The experimental setup includes a FTIR spectrometer (Nicolet Magna 550), in the spectra range from 4000 to 600 cm$^{-1}$ with a resolution of 1 cm$^{-1}$ in transmission mode. Sample analysis has been done in situ, just after each irradiation run. The 10:1 concentration of nitrogen:acetone vapor was obtained by imposing the related ratio partial pressure for both constituents in gas phase. Liquid acetone was purchased commercially from Sigma-Aldrich with purity greater than

99.8%. The preparation of acetone vapor was made after degassing through several freeze-pump-thaw cycles; N$_2$ was added in the preparatory chamber and the mixture was admitted into the analyzing chamber and condensed on a CsI substrate at 11 K.

The chamber pressure during the irradiation was below 2 x 10$^{-8}$ mbar, and IR spectra were acquired at different fluencies up to 1 x 10$^{13}$ ions cm$^{-2}$. Further details can be found in de Barros et al. (2011a). All the residual gases inside the chamber had a negligible partial pressure; the injected gaseous acetone-nitrogen mixture has condensed or was pumped out from the analysis chamber soon after closing the entrance gas valve; therefore, acetone layering was absent during irradiation. Common contaminants as well as the correspondence between laboratory conditions and astrophysics time scales have been discussed in details by de Barros et al. (2016).

The thickness of the ice mixture is calculated to be ~ 3.2 μm. This value is determined from the column densities of each precursor molecule and from their densities (0.83 g cm$^{-3}$ for pure solid nitrogen, Islam et al. (2014) and 0.79 g cm$^{-3}$, for liquid acetone, SRIM - Ziegler et al. (2010)); it is assumed that this thickness is preserved in a solid homogeneous ice mixture. For the current ice layer, its thickness is much shorter than the projectile range in the material, so that the destruction cross sections of both precursors can be considered approximately constant along the projectile track in the sample. This occurs because the energy loss of the projectile travelling inside the ice is negligible compared to its initial energy.

## 3 RESULTS

### 3.1 Precursor molecules

Fig. 1 shows the FTIR spectrum for the 10:1 N$_2$:CH$_3$COCH$_3$ mixture in the 3200-1000 cm$^{-1}$ range and Figure 2 compares segments of the mixture IR spectra: (i) as deposited (solid line) and (ii) after the end of irradiation (dotted line), corresponding to the fluence of 1 x 10$^{13}$ ions.cm$^{-2}$. The acetone features have been identified according to the work of Harris & Levin (1972) and Andrade et al. (2014). Fig. 3 displays the evolution of some IR spectrum segments of the mixture ice as a function of beam fluence.

Their wavenumbers, assignments and band strengths (referred in this work as A-values and expressed in cm molecule$^{-1}$) are listed in Table 1.

*(i) Acetone*

Sixteen acetone and one N$_2$ infrared (IR) peaks were observed. Since no A-value of the 1229.0 cm$^{-1}$ band was found in the literature for this ice mixture, we have used the same value quoted for pure acetone: $A_\nu^P$ = 1.27 x 10$^{-17}$ cm.molecule$^{-1}$. Figure 4 presents the acetone column density evolution determined from normalized absorbances of the fourteen acetone bands most visible in the spectrum at fluence 3 x 10$^{11}$ ions.cm$^{-2}$.

*(ii) Nitrogen*

For the N$_2$ molecule, only the 2328.4 cm$^{-1}$ band was identified based on the Sandford et al. (2001), Bernstein & Sandford (1999) and Jamieson et al. (2005) works. A discussion on A-values of the N$_2$ bands was performed by de Barros et al. (2015). The value $A_\nu^P$ = 2.38 x 10$^{-21}$ cm.molecule$^{-1}$for the 2328.4 cm$^{-1}$ band was calculated for the current 10:1 N$_2$:CH$_3$COCH$_3$ mixture, which is four orders of magnitude lower than the one for pure acetone (see Table 2).

Indeed, in a solid matrix, interactions between N$_2$ molecules with nearby polar molecules induce electric dipoles, which interact with IR photons creating N$_2$ vibrations; the N$_2$ molecular symmetry is broken and its vibrational modes become (weakly) active: the resulting small A-values remain however quite sensitive to the N$_2$ - acetone concentration (Sandford et al. 2001; Bernstein

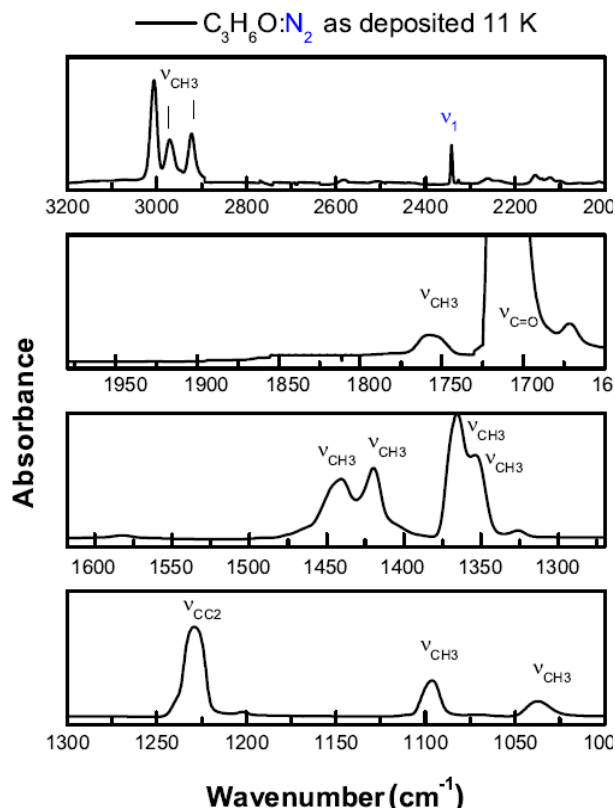

**Figure 1**. FTIR spectrum of the 3200 − 1000 cm$^{-1}$ range of the CH$_3$COCH$_3$:N$_2$ (1:10) ice at 11 K before irradiation.

& Sandford 1999).

### 3.1.1 Acetone and nitrogen destruction cross sections

Radiolysis of precursors in the ice mixture is studied by monitoring the absorbance for several bands of both precursors during the ion beam irradiation. Their destruction cross sections are obtained from the dependence of the respective molecular column densities on beam fluence. In turn, the column density N(F) is determined from the Lambert-Beer Law:

$$N(F) = \ln(10) \frac{S(F)}{A_\nu(F)} \quad (1)$$

where S(F) is the absorbance for a given band observed in the spectrum at beam fluence F. It has been observed that the A-value changes as the porosity or as the crystalline/amorphous state of the ice evolves with the fluence (Strazzulla et al. 1992; Dartois et al. 2015). Based on experimental data, Mejia et al. (2015) proposed

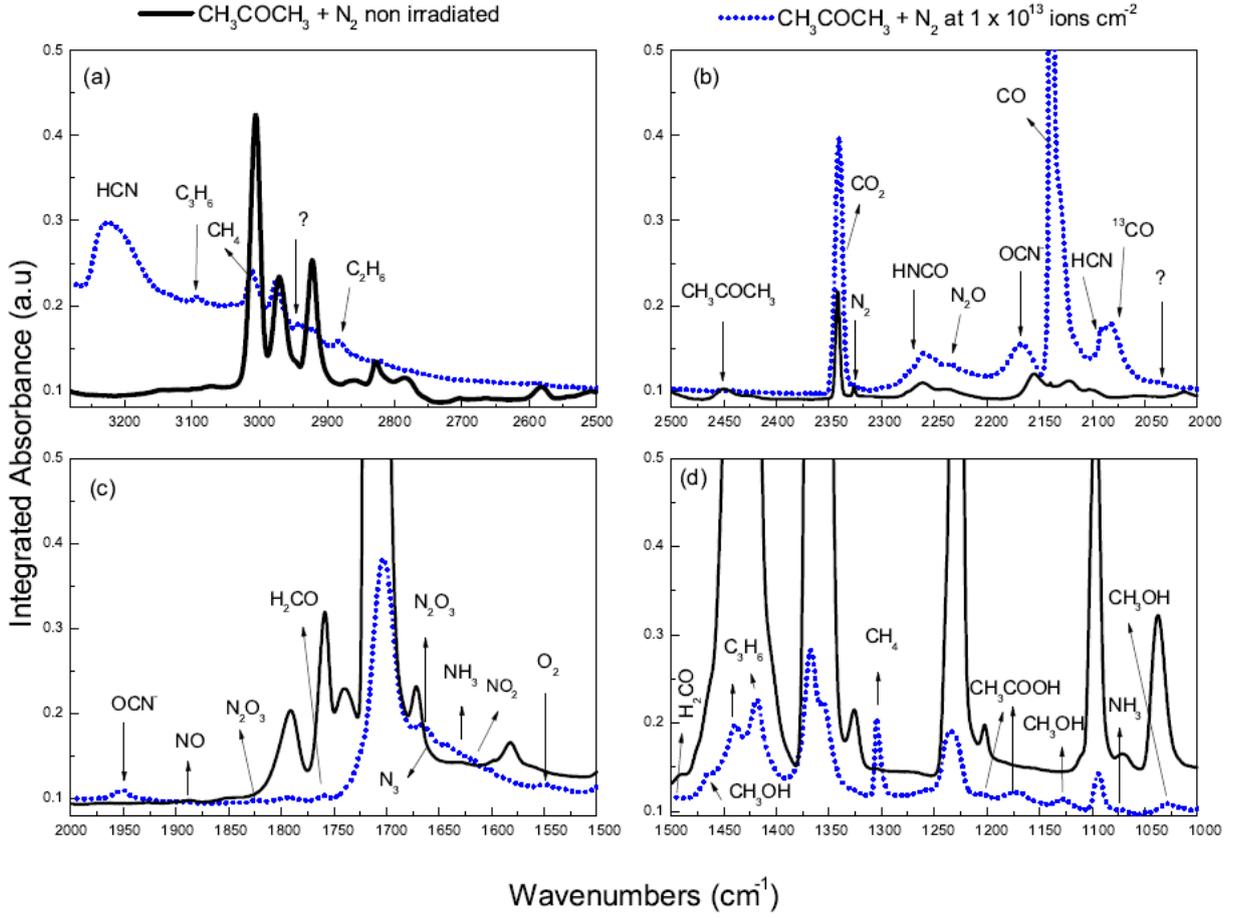

**Figure 2.** Experimental middle range IR spectra of N₂-acetone ice at 11 K, before (solid line) and after (dotted line) 40 MeV $^{58}$Ni$^{11+}$ irradiation with a fluence of 1 × 10$^{13}$ ions.cm$^{-2}$. Details of the bands at all irradiations are shown in Figure 3.

that $A_v(F)$ is described empirically by an exponential function:

$$A_v(F) = A_v^{eq}\left(1 - \zeta_{exp}(-\sigma_c F)\right) \quad (2)$$

where $A_v^{eq} = A_v(\infty)$ is the value after a enough long

$$S(F) = S_0[\exp(-\sigma_d^{ap} F) - \zeta \exp(-(\sigma_c + \sigma_d^{ap})F] \quad (3)$$

irradiation (situation in which the ice crystalline states are brought into equilibrium by the projectiles), the parameter $\zeta = (S_0 - S(0))/S_0$ is the initial relative porosity and $\sigma_c$ is the compaction cross section. Here, S(0) is the absorbance for a given band, measured for the virgin ice (F = 0) as deposited; $S_0$ corresponds to the initial absorbance for a previously compacted virgin ice, or, alternatively, it can be determined by extrapolation at F = 0 from $A_v(F)$ data at high fluences.

These considerations are relevant because the A-values reported in literature usually correspond to the ice as deposited (F = 0) which are often porous, therefore related to S(0) and not to $S_0$. For clearness, the A-value for the porous ice is called here $A_v^p$, which is equal to $A_v(0)$ and connected with the one for equilibrium states by $A_v^p = A_v^{eq}(1 - \zeta)$, according to Eq. (2).

Irradiation changes the IR absorbance, S(F), through three processes: (i) compaction, (ii) radiolysis and (iii) sputtering.

Mejia et al. (2015) and de Barros et al. (2015) suggested that, in the absence of continuous deposition (layering) during the measurements, the evolution of S(F) due to these three processes is given by the empirical function:

where $\sigma_d^{ap}$ (or $\sigma_{d,i}^{ap}$, $i$ being the precursor label) is an apparent destruction cross section which parameterizes the simultaneous effects of radiolysis and sputtering processes. From Eqs. (1) to (3), the expected column density evolution of the precursor molecules is determined by:

$$N(F) = N_0 \exp(-\sigma_d^{ap} F) \quad (4)$$

where

$$N_0 = \ln(10)\frac{S_0}{A_v^{eq}} = \ln(10)\frac{S(0)}{A_v^p} \quad (5)$$

Briefly, the current analysis consists in fitting the absorbance evolution of each of the precursor molecule bands with Eq. (3) in order to extract $S_0$, $\zeta$, $\sigma_c$ and $\sigma_d^{ap}$. Using S(0) = $S_0$ (1- $\zeta$), the available literature value of $A_v^p$ and Eq. (5), the initial column density, $N_0$, of that precursor species is determined.

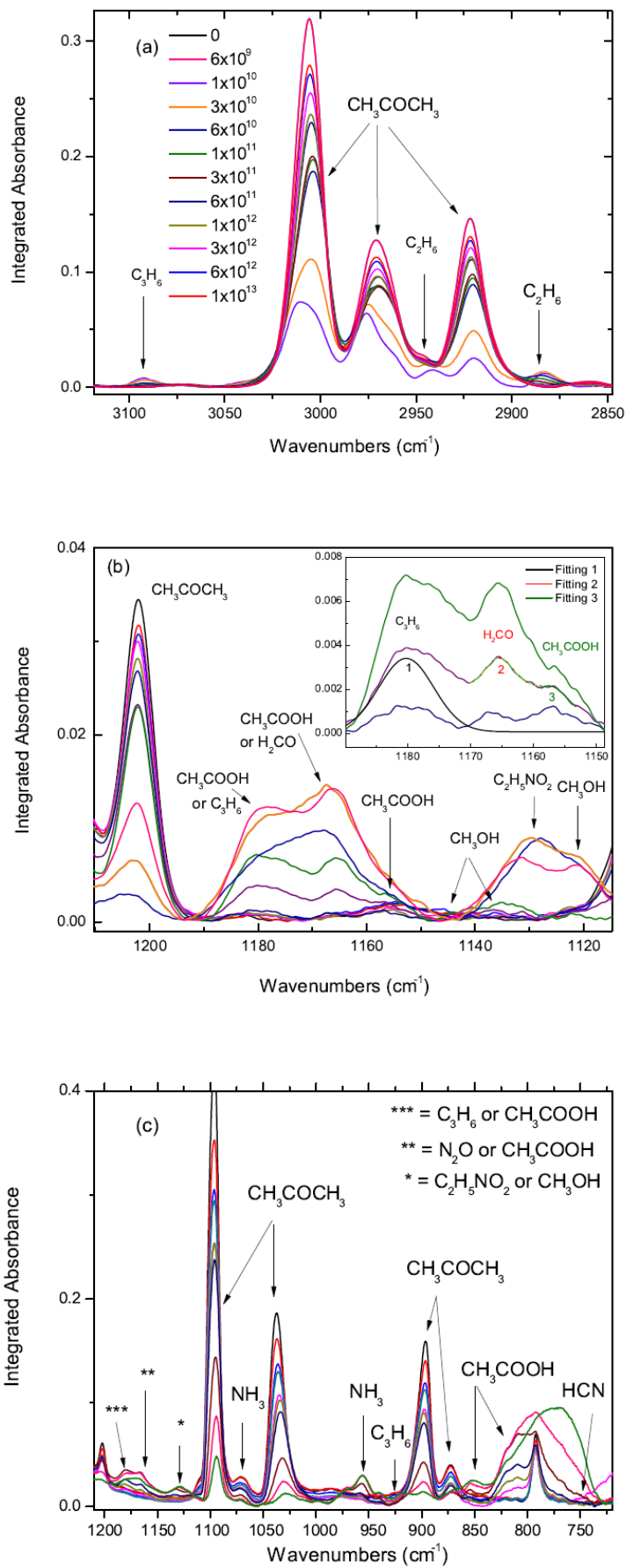

**Figure 3**. FTIR spectra in (a) 3200 - 2850 cm$^{-1}$ range, (b) 1250 - 1100 cm$^{-1}$ range (the inset correspond to a gausssian fittings to the molecules: 1 = $C_3H_6$ or $CH_3COOH$, 2 = H2CO or $CH_3COOH$ and 3 = $CH_3COOH$) and (c) 1230 - 700 cm$^{-1}$ range of the $CH_3COCH_3$:$N_2$ (1:10) ice at 11 K during irradiation.

Table 1. CH$_3$COCH$_3$ and N$_2$ wavenumbers and vibration modes observed in this work and their respective band strength A-values ($A_v^p$) before irradiation for the porous ice mixture and for the pure acetone porous ice.

| Wavenumber (cm$^{-1}$) | Assignments | A-values for the mixture ($\times 10^{-17}$ cm molecule$^{-1}$) | A-values for precursor ($\times 10^{-17}$ cm molecule$^{-1}$) |
|---|---|---|---|
| CH$_3$COCH$_3$ | | mixture | pure |
| 791.8 | CC$_2$ sym. stretch[a] | 0.067 ± 0.001 | - |
| 871.3 | CH$_3$ rocking mode[a] | 0.045 ± 0.001 | 0.020 ± 0.001 |
| 897.9 | CH$_3$ rocking mode[a] | 0.07 ± 0.01 | 0.050 ± 0.003 |
| 1036.7 | CH$_3$ anti-sym. deformation[c] | 0.27 ± 0.01 | - |
| 1073.7 | CH$_3$ rocking mode[a] | 0.032 ± 0.008 | 0.017 ± 0.004 |
| 1096.8 | CH$_3$ rocking mode[a] | 0.41 ± 0.01 | 0.21 ± 0.06 |
| 1229.0 | CC$_2$ anti-sym. stretch[a,b] | 1.27* | 1.27 |
| 1352.8 | CH$_3$ sym. deformation[a] | 1.0 ± 0.3 | 0.50 ± 0.03 |
| 1366.2 | CH$_3$ sym. deformation[a] | 1.5 ± 0.6 | 1.3 ± 0.4 |
| 1419.7 | CH$_3$ sym. deformation[a] | 0.65 ± 0.04 | 0.45 ± 0.06 |
| 1442.5 | CH$_3$ sym. deformation[a] | 1.2 ± 0.4 | 0.76 ± 0.06 |
| 1710.9 | C=O stretch[a] | 6.4 ± 0.9 | 5.2 ± 1.0 |
| 1758.7 | - | 0.18 ± 0.03 | 0.030 ± 0.002 |
| 2921.8 | CH$_3$ asym. stretch[a] | 0.26 ± 0.03 | 0.10 ± 0.03 |
| 2970.2 | CH$_3$ asym. stretch[a] | 0.29 ± 0.03 | 0.12 ± 0.03 |
| 3006.3 | CH$_3$ asym. stretch[a] | 0.50 ± 0.03 | 0.24 ± 0.03 |
| N$_2$ | | A-value | |
| 2328.4 | N-N stretch[c] | 2.38 × 10$^{-4}$ * | - |

[a]Andrade et al. (2014) (for pure acetone ice), [b]Harris & Levin (1972), [c]This work. *A-values used in the cross section calculations.

For the acetone cross section calculations, we choose the acetone skeletal (C-C$_2$ vibration mode) at 1229.0 cm$^{-1}$ as reference band. The compaction cross section of (1.64 ± 0.30) x 10$^{-11}$ cm$^2$ and the destruction cross section of (2.52 ± 0.60) x 10$^{-13}$ cm$^2$ are obtained (Fig. 5 and Table 2).

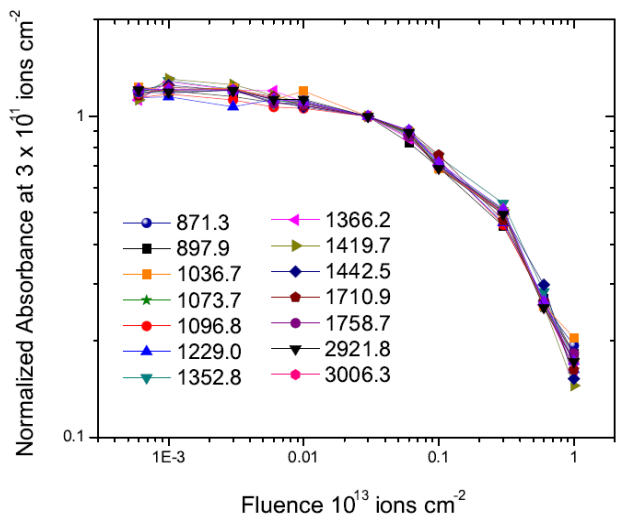

**Figure 4.** Evolution of the acetone molecular absorbance as a function of the projectile fluence. The different lines correspond to the different vibration modes observed in the IR spectra (Table 1). For better comparison, the band absorbance is normalized to the 1229 cm$^{-1}$ band one at 3 × 10$^{11}$ ions cm$^{-2}$ fluence.

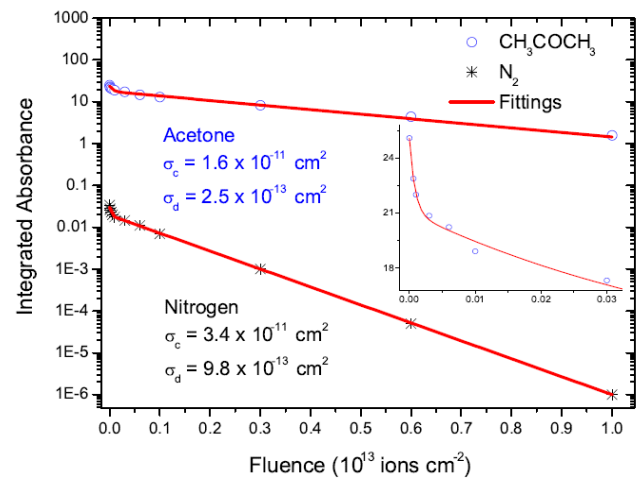

**Figure 5.** Absorbance of the 2328 cm$^{-1}$ nitrogen and 1229 cm$^{-1}$ acetone bands as a function of the beam fluence. In the inset, a zoom on the acetone ice degradation for low fluences. The obtained fitting parameters, using Eq. (3), are presented in Table 2.

The 2328.4 cm$^{-1}$ band N-N stretch is used as reference to determine the nitrogen compaction and destruction cross section, as can be seen in Figure 5. For the N$^2$ precursor, the compaction cross section obtained is (3.4 ± 0.5) x 10$^{-11}$ cm$^2$ and the destruction one is (9.8 ± 0.7) x 10$^{-13}$ cm$^2$. Since the nitrogen molecules are destroyed faster than acetone molecules, a relatively large number of free N atoms is delivered into the ice since the beginning of irradiation; indeed, for the highest fluences, the $N_2$ peaks practically disappear and just the acetone ones remain in the spectra.

It is known that A-values may also vary with the precursor concentration, particularly for homoatomic molecules such as $N_2$; for $N_2$:$H_2O$ mixture, Sandford et al. (2001) found that this variation follows a power law, rather than an exponential one (de Barros et al. 2015): lower the $N_2$ relative concentration, higher its A-value is. The calculated $N_2$ column density should then decay faster than an exponential function. For the present calculation, however, we have considered that the $N_2$ A-value remained constant during irradiation: its column density behavior should therefore be seen with caution.

Table 2. Parameters obtained in the acetone and nitrogen data fitting with Eq. 3. $N_0 = \ln(10)\, S_p/A_\nu^p$, where $A_\nu^p$ is given in the literature. The $N_2$ $A_\nu^{eq}$ was assumed to remain constant during the irradiation. The negative value of $\zeta$ means that the ice compaction reduces the IR absorbance ($S_0 < S_p$).

|  | $CH_3COCH_3$ | $N_2$ |
|---|---|---|
| Reference band (cm$^{-1}$) | 1229.0 | 2328.4 |
| $S_p$ | 2.75 | 0.03 |
| $S_0$ | 2.02 | 0.02 |
| $\zeta = (S_0 - S_p)/S_0$ | −0.36 | −0.48 |
| $\sigma_c$ ($10^{-11}$ cm$^2$) | 1.64 | 3.4 |
| $\sigma_d^{ap}$ ($10^{-13}$ cm$^2$) | 2.52 | 9.8 |
| $A_\nu^p$ ($10^{-17}$ cm molecule$^{-1}$) | 1.27 | 0.000 24 |
| $A_\nu^{eq} = A_\nu^p (1-\zeta)^{-1}$ | 0.74 | 0.000 13 |
| $N_0$ ($10^{18}$ molecule cm$^{-2}$) | 0.65 | 32.9 |

### 3.2 Daughter Molecules

Seventeen daughter species have been observed after the Ni ion bombardment of the acetone-nitrogen ice mixture: CO, $CO_2$, $CH_4$, $C_2H_6$, $C_3H_6$, $CH_3OH$, $H_2CO$, $CH_3COOH$, NO, $N_2O_3$, $N_3$, $NO_2$, HNCO, HCN, $NH_3$, OCN, $C_2H_5NO_2$.

All the observed daughter species may be classified as: (i) direct products of acetone only (in the sense that they do not have N atoms), (ii) direct product of nitrogen (only the $N_3$ species) and (iii) hybrid species (daughters of both precursors). Table 3 displays the A-values used for the formation cross section calculations of the daughter species.

#### 3.2.1 Acetone products

##### CO and $CO_2$

Two carbon monoxide absorption features are observed: the $\nu_1$ band, characterized by a relatively narrow peak around 2139 cm$^{-1}$, and the 2$\nu_1$ band at 4251 cm$^{-1}$ (Jamieson et al. 2006). The carbon dioxide is identified through the intense CO stretch ($\nu_3$) band at 2348 cm$^{-1}$ (Jamieson et al. 2006).

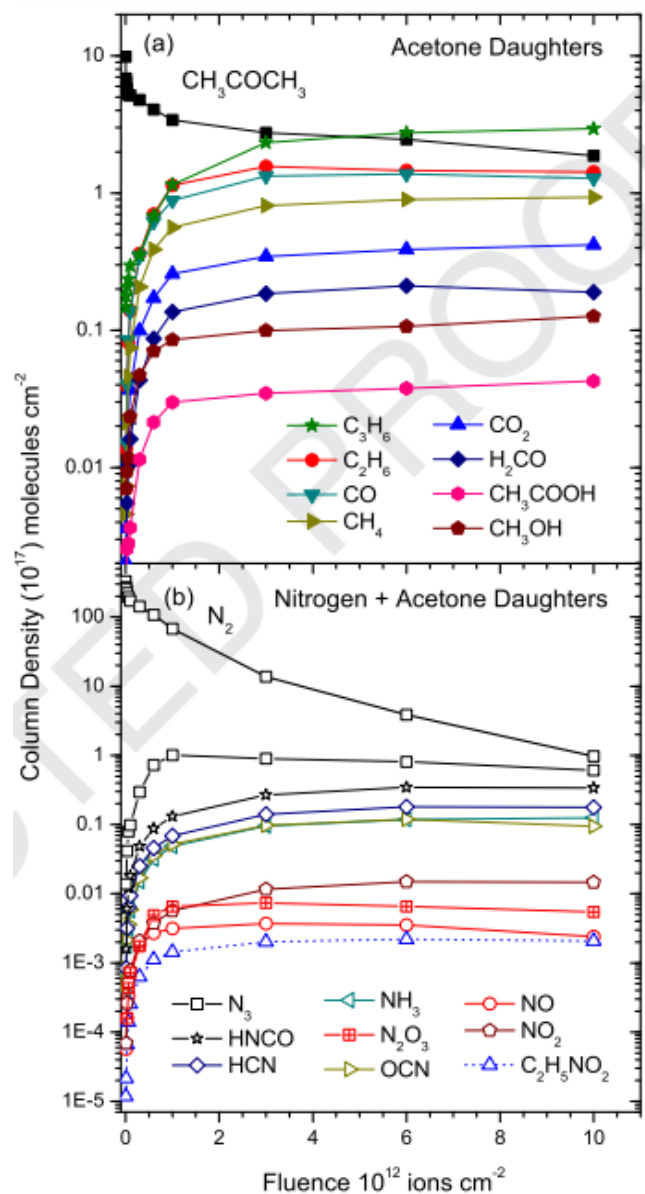

**Figure 6.** Column density evolution for daughter species produced by the $CH_3COCH_3$:$N_2$ radiolysis. (a) $CH_3COCH_3$ fragmentation: CO, $CO_2$, $CH_4$, $C_2H_6$, $C_3H_6$, $CH_3OH$, $H_2CO$ and $CH_3COOH$ synthesis; (b) $N_3$ from the $N_2$ radiolysis and hybrid species $N_2$ fragmentation reactions: $CH_3COCH_3$ + $N_2$: OCN, NO, $N_2O_3$, HNCO, HCN, $C_2H_5NO_2$, $NH_3$ and $NO_2$ synthesis. Solid curves are guides for the eyes.

##### $CH_4$, $C_2H_6$ and $C_3H_6$

Two $CH_4$ bands are present: its $\nu_4$ band at 1303 cm$^{-1}$, the one located at 4284 cm$^{-1}$ (Bohn et al. 1994; Moore & Hudson 1998; Gerakines et al. 2005; de Barros et al. 2011a) are seen. Unfortunately, its $\nu_1$ symmetric stretch band at 2917 cm$^{-1}$ and its $\nu_3$ stretch band at 3009 cm$^{-1}$ (see Fig. 2(a)) are overlapped by the 2922 and 3006 cm$^{-1}$ acetone bands (de Barros et al. 2011a; Mejia et al. 2015).

Three ethane bands located at 2975, 2882 and 1463 cm$^{-1}$ are observed (Bohn et al. 1994; Moore & Hudson 1998; de Barros et al. 2011a, 2017). The propene is observed through five bands: 3092, 1438, 1420, 1174 and 935 cm$^{-1}$ (Fig. 3(a)) and the intense $\nu_1$ band ($CH_2$ asymmetric stretch) at 3092 cm$^{-1}$ and $\nu_{11}$ band at 1174 cm$^{-1}$ (Figs. 3(b)-inset and 3(c)) as well seen. The 1438 and 1420 cm$^{-1}$ bands are too close to the acetone's 1440 and 1419.7 cm$^{-1}$ ones. The $\nu_9$ 1438 cm$^{-1}$ ($CH_2$ scissor) and the $\nu_{16}$ at 1442 cm$^{-1}$ ($CH_3$ asymmetric deform) are also shown in details in Figures 2(d) (Hiraoka et al. 2002; NIST 2011).

Table 3. Molecular species, wavenumber and A-value corresponding to the band used to calculate the cross sections.

| Species | Wavenumber (cm$^{-1}$) | A-value (10$^{-17}$ cm molecule$^{-1}$) | Reference |
|---|---|---|---|
| CO | 4251, 2139$^a$ | 1.1 | Gerakines et al. (1995); Jamieson et al. (2006) |
| CO$_2$ | 2348$^a$ | 7.6 | Gerakines et al. (1995); Jamieson et al. (2006) |
| CH$_4$ | 3009, 2917, 1303$^a$ | 0.78 | de Barros et al. (2011a); Mejia et al. (2015) |
| C$_2$H$_6$ | 2975, 2882$^a$, 1463 | 0.32 | de Barros et al. (2016); Moore & Hudson (1998) |
| C$_3$H$_6$ | 3092$^a$, 1438, 1420, 1174, 937 | 0.078$^b$ | NIST (2011) |
| CH$_3$OH | 2830, 1460$^a$, 1129, 1074, 1029 | 0.91 | de Barros et al. (2011b) |
| CH$_3$COOH | 2244, 1181, 1166$^a$, 847 | 1.96 | de Barros et al. (2011b) |
| H$_2$CO | 2825, 1758$^a$, 1497, 1170 | 0.96 | Bohn et al. (1994); de Barros et al. (2011a) |
| O$_2$ | 2139, 1550 | – | de Barros et al. (2014) |
| N$_3$ | 1657, 1653$^a$, 1160 | 0.23 | Jamieson et al. (2005) |
| NH$_3$ | 1070$^a$, 1627, 950 | 1.7 | Sandford & Allamandola (1993) |
| NO | 1875 | 0.68/0.45 | Jamieson et al. (2005); Sicilia et al. (2012) |
| NO$_2$ | 1616 | 6.36 | Jamieson et al. (2005); de Barros et al. (2016) |
| N$_2$O$_3$ | 1834$^a$, 1654, 1304 | 6.4 | Jamieson et al. (2005); de Barros et al. (2017) |
| HNCO | 2260$^a$ | 7.2 | Kanuchova et al. (2016) |
| HCN | 3275, 2096$^a$, 747 | 1.1 | Moore & Hudson (2003) |
| OCN | 2165, 1948$^a$ | 1.0 | Sicilia et al. (2012); Moore & Hudson (2003) |
| C$_2$H$_5$NO$_2$$^c$ | 1600, 1505, 1334, 1130, 1112, 1034$^a$ | 0.14 | Portugal et al. (2014) |

$^a$A-values listed in the third column, which have been used for the cross-section calculations; $^b$this work; $^c$not taken into account in our analysis.

### CH$_3$OH

Five methanol bands are identified at: 2830 cm$^{-1}$ ($v_3$ - C-H parallel symmetric stretch), 1460 cm$^{-1}$ ($v_4$ C-H in-plane asymmetric band), 1129 cm$^{-1}$ ($v_{11}$ CH$_3$ rock), 1074 cm$^{-1}$ ($v_7$ CH$_3$ rock) and 1029 cm$^{-1}$ ($v_8$ C-O stretch) (de Barros et al. 2011b). Since the 1074 cm$^{-1}$ one is close to an acetone band and the 1129 and 1029 cm$^{-1}$ ones are close to possible glycine bands, the $v_4$ (1460 cm-1) feature was chosen for the methanol cross section calculations.

### CH$_3$COOH

Four acetic acid bands are seen at 2244, 1181, 1166 and 847 cm$^{-1}$, respectively. The $v_{13}$ (2996 cm$^{-1}$ - CH3 d-stretch) ones are not seen because of the overlapping with the 3006 cm$^{-1}$ acetone band (see Fig. 3(a)). The $v_8$ C-O stretch at 1181 cm$^{-1}$ can also be identified as C$_3$H$_6$ (Fig. 3(b)). The $v_{10}$ C-C stretch at 847 cm$^{-1}$ band is close to acetone bands but is enough visible to be analyzed as shown in Fig. 3(c). The 1166 cm$^{-1}$ band was used for the acetic acid cross section calculations (de Barros et al. 2011b).

### H$_2$CO

The formaldehyde has four main bands which are located at 2825, 1758, 1497 and 1170 cm$^{-1}$, respectively (Bohn et al. 1994; de Barros et al. 2011b). The $v_1$ 2825 cm$^{-1}$, CH$_2$ stretch band, is close to the 2830 cm$^{-1}$ methanol band. The $v_3$ 1497 cm$^{-1}$, CH$_2$ scissor, is close to the 1505 cm$^{-1}$ glycine band. The $v_4$ at 1170 cm$^{-1}$ is too close to the one at 1166 cm$^{-1}$ due to N$_2$O and/or CH$_3$COOH. The only reliable formaldehyde band is the $v_2$ 1758 cm$^{-1}$, a CO stretch band: even being close to an acetone feature, this band is well defined (Fig. 2(c)) and was used for the formaldehyde cross section calculations.

### O$_2$, O$_3$

Unexpected peaks due to O$_2$ at 1550 cm$^{-1}$ and 2139 cm$^{-1}$ are also observed in a high fluence irradiation (de Barros et al. 2014). The 2139 cm$^{-1}$ one is very close to the $v_1$ band of carbon monoxide, but the 1550 cm$^{-1}$ band can be well identified in the current experiment (Fig.2(c)). In principle O$_3$ could be formed in the acetone radiolysis, but its 1042 cm$^{-1}$ band cannot be observed because it is too close to the 1036.7 cm$^{-1}$ acetone band.

### 3.2.2 Nitrogen products

### N$_3$

N$_2$ is the dominant molecule in the current mixture. Its radiolysis delivers large quantity of atomic nitrogen in the matrix, so that N$_3$ formation is expected. The known N$_3$ IR bands are: 1160, 1653 and 1657 cm$^{-1}$ (Hudson & Moore 2002; Jamieson & Kaiser 2007). The wavenumber of the most intense ($v_3$ mode) coincides with that of the N$_2$O$_3$ $v_2$ mode (Jamieson et al. 2005) and is very close of the NH$_3$ 1627 cm$^{-1}$ (see Fig.2(c)) and of the NO$_2$ 1613 cm$^{-1}$ bands (Jamieson et al. 2005; Sandford & Allamandola 1993). As discussed in the Section 3.2.3, the 1654 cm$^{-1}$ band is a relatively low intense N$_2$O$_3$ band, but the NH$_3$ 1627 cm$^{-1}$ and NO$_2$ 1613 cm$^{-1}$ ones are stronger.

### 3.2.3 Nitrogen and acetone products

### N$_2$O

Two nitrous oxide bands should be considered: 1295 and 1166 cm$^{-1}$. The band at 1166 cm$^{-1}$ (2 $v_2$) coincides with the 1166 cm$^{-1}$ CH$_3$COOH band and the $v_1$ one is a very small feature at 1295 cm$^{-1}$ close to the CH$_4$ 1300 cm$^{-1}$ band (Fulvio et al. 2009; de Barros et al. 2015). Therefore we are not using this molecule for calculations.

### NO$_2$ and NO

The $v_3$ at 1616 cm$^{-1}$ band can be attributed to nitrogen dioxide (Jamieson et al. 2005; de Barros et al. 2016) and the $v_1$ at 1875 cm$^{-1}$ to nitric oxide (Sicilia et al. 2012). Since both bands are very intense, they have been used for the NO$_2$ and NO column density calculations, respectively. Formation of nitrogen oxides by radiolysis has been discussed recently (Almeida et al. 2017).

Table 4. Molecular species, formation ($\sigma_f$), effective destruction ($\sigma_d^{eff}$) and destruction ($\sigma_d$) cross sections obtained by fitting the column density evolutions of daughter species with Eq. (7).

| Species | $\sigma_f$ (this work) ($\times 10^{-13}$ cm$^2$) | $\sigma_f$ (pure*) ($\times 10^{-13}$ cm$^2$) | $\sigma_d^{eff}$ (this work) ($\times 10^{-13}$ cm$^2$) | $\sigma_d$ (this work) ($\times 10^{-13}$ cm$^2$) | $\sigma_d$ (pure*) ($\times 10^{-13}$ cm$^2$) |
|---|---|---|---|---|---|
| CO | 0.41 ± 0.07 | 0.15 ± 0.05 | 1.9 ± 0.4 | 0.62 ± 0.06 | 0.016 ± 0.004 |
| CO$_2$† | 0.32 ± 0.08 | 0.26 ± 0.06 | 1.6 ± 0.5 | 0.80 ± 0.08 | 0.035 ± 0.001 |
| CH$_4$ | 0.22 ± 0.06 | 0.66 ± 0.09 | 1.8 ± 0.5 | 0.69 ± 0.09 | 0.16 ± 0.05 |
| C$_2$H$_6$ | 0.63 ± 0.08 | 0.060 ± 0.003 | 1.9 ± 0.4 | 0.62 ± 0.06 | 0.24 ± 0.06 |
| C$_3$H$_6$ | 1.75 ± 0.57 | - | 2.1 ± 0.8 | 0.42 ± 0.09 | - |
| CH$_3$OH | 0.035 ± 0.009 | - | 1.9 ± 0.7 | 0.67 ± 0.08 | - |
| CH$_3$COOH† | 0.028 ± 0.005 | - | 1.6 ± 0.3 | 0.89 ± 0.08 | - |
| H$_2$CO | 0.14 ± 0.09 | 0.40 ± 0.02 | 1.8 ± 0.5 | 0.71 ± 0.06 | 0.14 ± 0.04 |
| N$_3$† | 0.35 ± 0.06 | - | 7.7 ± 0.8 | 5.21 ± 0.06 | - |
| NH$_3$ | 0.0035 ± 0.0006 | - | 1.9 ± 0.7 | 0.60 ± 0.07 | - |
| NO | 0.00052 ± 0.00009 | - | 2.0 ± 0.5 | 0.51 ± 0.06 | - |
| NO$_2$† | 0.0063 ± 0.0009 | - | 1.9 ± 0.7 | 0.60 ± 0.06 | - |
| N$_2$O$_3$† | 0.022 ± 0.009 | - | 9.3 ± 0.6 | 6.8 ± 0.06 | - |
| HNCO | 0.0092 ± 0.0007 | - | 7.3 ± 0.6 | 4.78 ± 0.08 | - |
| HCN | 0.0024 ± 0.0003 | - | 1.4 ± 0.4 | 1.13 ± 0.06 | - |
| OCN | 0.0037 ± 0.0007 | - | 1.9 ± 0.6 | 0.60 ± 0.08 | - |
| C$_2$H$_5$NO$_2$‡ | 0.00023 ± 0.00007 | - | 1.6 ± 0.5) | 0.91 ± 0.09 | - |

* for pure acetone (Andrade et al. 2014);
†Cross sections are multiplied by two/three, because these species need two/three precursor molecules to be formed
‡Not considered in the calculation.

### N$_2$O$_3$

Three di-nitrogen tri-oxide bands are observed. One relatively small, at 1654 cm$^{-1}$ ($v_2$, NO$_2$ - asymmetrical stretch), one at 1304 cm$^{-1}$ ($v_3$, NO$_2$ symmetrical stretch), and another at 1834 cm$^{-1}$ ($v_1$, N=0 - stretch) (Jamieson et al. 2005; de Barros et al. 2017). The 1304 cm$^{-1}$ band is close to a CH$_4$ band, and the one located at 1654 cm$^{-1}$ is close to a NH$_3$ band. Therefore the band located at 1834 cm$^{-1}$ was used for cross section and budget calculations.

### NH$_3$

Three ammonia bands are observed (Sandford & Allamandola 1993). The $v_4$ deformation at 1627 cm$^{-1}$, the 950 cm$^{-1}$ and the 1070 cm$^{-1}$ ones, as shown at Fig.3(c). The $v_2$ deformation band at 1070 cm$^{-1}$ was used for the cross section calculation.

### OCN

Two isocyanato radical bands are observed at 2165 and 1948 cm$^{-1}$ (Sicilia et al. 2012; Portugal et al. 2014). Both bands can be perfectly seen in Fig.2(b). The $\Sigma^{+3}$ asym. stretch at 1948 cm$^{-1}$ feature was used for the calculations (NIST 2011). We assume that the 1942 cm$^{-1}$ band observed by Sicilia et al. (2012) is blueshifted towards the 1948 cm$^{-1}$ position in the present measurements. (Fig.2(c)).

### HCN

Three hydrogen cyanide bands are observed in the current experiment. They are located at 3275 cm$^{-1}$ ($v_3$ - CH stretch), 2096 cm$^{-1}$ ($v_1$ - CN stretch) and possibly 747 cm$^{-1}$ ($v_2$ - bend). (Moore & Hudson 2003). For high fluence irradiation, the band located at position 3275 cm$^{-1}$ may be blueshifted towards the 3245 cm$^{-1}$ position, as can be seen in Fig.2(a). Moore & Hudson (2003) reported that the band at 2096 cm$^{-1}$ may be identified as CH$_2$N$_2$.

### HNCO

An isocyanic acid band, corresponding to the H-N=C=O a-stretch vibration at 2260.0 cm-1 (Kanuchova et al. 2016), is clearly observed (Fig.2(b)) and has been taken into account in the analysis.

### C$_2$H$_5$NO$_2$

Finally, there are six candidate bands for the glycine. They are located at: 1600 cm$^{-1}$ ($v_{as}$ - $CO_2^-$ asymmetric stretch), 1505 cm$^{-1}$ ($\delta_s$ NH$_3$ scissoring), 1335 cm$^{-1}$ (w CH$_2$ wagging), 1130 cm$^{-1}$ (NH$_3$ stretching), 1112 cm$^{-1}$ ($\rho$ NH$_3$ rocking- Fig.3(c)) and 1034 cm$^{-1}$ ($v$CN stretching) (Portugal et al. 2014). The 1034, 1112, 1335 (Fig.2(d)), 1413 and 1505 cm$^{-1}$ bands are very close to acetone bands and their identification or analysis is not reliable. The 1130 cm-1 band (Fig.3(b)) is too close to the 1129 cm$^{-1}$ methanol band (de Barros et al. 2011b). Near 1600 cm$^{-1}$, daughter features appear with irradiation, but resolution is not good enough for discrimination.

Since all the bands that may be due to glycine have ambiguous identification, we are not taking its formation into account in the budget calculations nor extracting its cross section. Nevertheless, due to the relevance of glycine, in Section **5.3** a short discussion on its **possible** formation is presented.

### 3.2.4 Daughters molecules cross section calculations

The A-values of some bands of thirteen nitrogen-acetone daughter species are known and, therefore, their column densities can be calculated. Figs.6(a) and (b) show the column density evolution for the observed direct and hybrid products, respectively, as a function of the fluence. Figs 5 and 6(b) show that the N$_2$ column density decays very fast. This behavior can be explained by the

fact that although N₂ molecules disappear quickly during irradiation, carbon, hydrogen and oxygen atoms from

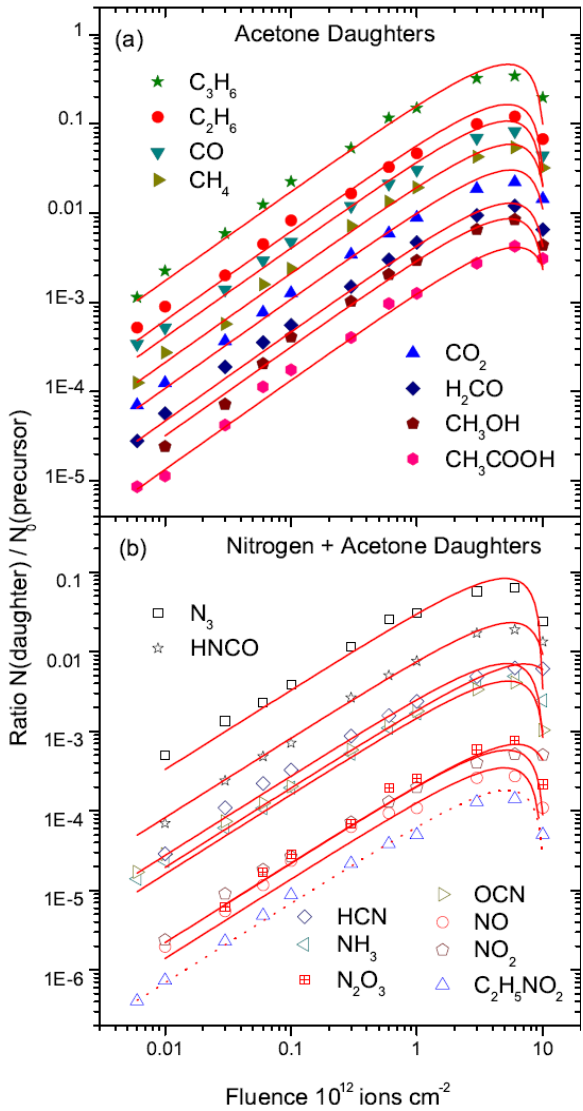

**Figure 7.** Evolution of the column densities of daughter species produced in the $CH_3COCH_3:N_2$ radiolysis. (a) $N/N_0$ ratio for the direct products of $CH_3COCH_3$: CO, $CO_2$, $CH_4$, $C_2H_4$, $C_2H_6$, $C_3H_6$, $CH_3OH$, $CH_2OH$, $H_2CO$ and $CH_3COOH$; (b) $N/N_0$ ratio for the $N_3$ from the $N_2$ radiolysis and hybrid species $N_2$ fragmentation reactions: $CH_3COCH_3 + N_2$: OCN, NO, $N_2O_3$, HNCO, HCN, $C_2H_5NO_2$, $NH_3$ and $NO_2$ synthesis. Solid curves are Eq. (4) predictions. The extracted cross sections are presented in Table 3.

acetone may react with atomic nitrogen to form hybrid molecules. Daughter species produced from $CH_3COCH_3:N_2$ ice mixture irradiation are listed in Table 3.

The formation ($\sigma_{f,j}$) and destruction ($\sigma_{d,j}$) cross sections for the daughter species ($\sigma_j$) were calculated by the kinetic model based on the evolution of the column density $N_i(F)$ (de Barros et al. 2016):

$$\frac{dN_j(F)}{N_{i,0}} = \frac{\sigma_{f,j}}{(\sigma_{d,i}^{ap} - \sigma_{d,j}^{ap})}\left[\exp\left(-\sigma_{d,j}^{ap}F\right) - \exp(-\sigma_{d,i}^{ap}F)\right] \quad (6)$$

Hybrid daughter molecules are products of two or more precursors. For these species, the index $i$ specifies the ensemble of precursor species required to form a given daughter species $j$. Eq. (6) is still valid for hybrid molecules but the precursor column density $N_{i,0}$ should be interpreted as the column density of the precursor ensemble $i$. The lowest concentration precursor ($N_{<,0}$) should play the dominant role in the chemical reaction velocities, in the sense that its exhaustion stops the daughter formation.

Nevertheless, it is important to note that the measured cross sections are correct only for the used relative concentrations of precursors. For low fluence irradiation, the concept of effective destruction (destruction of both, precursor and product) cross section may be invoked once the parameter $\sigma_d^{eff} = \sigma_{d,i}^{ap} + \sigma_{d,j}^{ap}$ is the one directly determined from the fitting procedures. Indeed, for $\sigma_d^{eff} F \ll 1$, Eq. (6) can be expanded into:

$$\frac{N_j(F)}{N_{<,0}} \sim (\sigma_{f,j}F)\left[1 - \frac{1}{2}\sigma_d^{eff}F\right] \quad (7)$$

Keeping this value fixed, Eq. (7) is used to fit daughter species data. From Eq. (3), the value $\sigma_{d,i}^{ap}$ for acetone found to be $(2.5 \pm 0.6) \times 10^{-13}$ cm². Figures 7(a) and 7(b) show the fittings for the formed species; the cross sections obtained by this procedure are listed in Table 4. In this Table, the first column displays the observed daughter species, the second column shows the formation cross section for species produced in the present ice mixture, and the third one presents the formation cross section for those produced in the pure acetone ice (Andrade et al. 2014). The three last columns show, respectively: the effective destruction cross section for the formed species in the ice mixture sample, the destruction cross sections for the same species, obtained with Eq. (7); the destruction cross sections for species formed from pure acetone ice (Andrade et al. 2014).

## 4 ATOM BUDGET

The issue of the budget is addressed by testing whether the column density decrease, ΔN, of the destroyed acetone and nitrogen is compatible with the column density increase of the products that have been observed by FTIR. Considering $F_{end} = 1.00 \times 10^{13}$ ions.cm⁻² as the final fluence, it is found that $\Delta N = N(F_{end}) - N_0 = 5.5 \times 10^{17}$ acetone and $3.29 \times 10^{19}$ nitrogen molecules cm⁻² have been destroyed (Table 5).

The molecular and atomic yields over the irradiation, defined as the average number of molecules or atoms per projectile that have been submitted to chemical reaction or sputtering, are determined using the calculated column density variations of the precursor and of the formed species. As shown in Table 5, the yield of $33 \times 10^4$ hydrogen atoms, $16.5 \times 10^4$ carbon atoms, and $5.5 \times 10^4$ oxygen atoms (the constituents) are removed from the acetone molecules per each projectile-ice collision. For $N_2 \sim 6.6 \times 10^6$ free nitrogen atoms per projectile are delivered into the ice, fraction of them transformed into the nitrogen-content of daughter species.

Comparing these yields with the formation rate of all the sixteen daughter species, 84% of the oxygen, the majority of hydrogen (~ 96%) atoms, 56% of carbon atoms and 0.44% of nitrogen atoms are observed in the products.

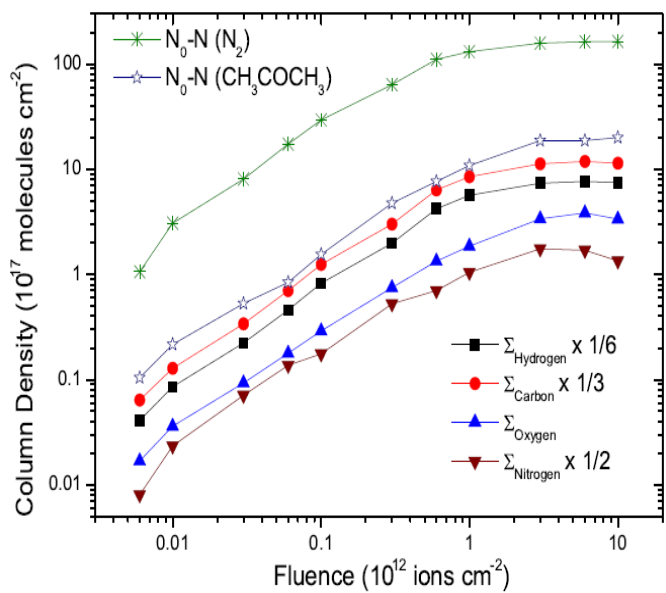

**Figure 8.** Column density variations of the two precursor molecules compared with the sum of the column density of constituent atoms (divided by the number of atoms forming the precursors).

As a result, if the A-values reported in Tables 1 and 3 are correct, we observe by FTIR practically all the molecules that contain hydrogen in their composition, in contrast with oxygen, carbon and nitrogen molecules. Most of the missing atomic species have probably desorbed and very likely are $C_nH_mN_q$ molecules, e.g., $NH_2CH_2$, $CH_2NH_2$ and the carboxyl radical, one of the species abundantly produced from the acid acetic photo-dissociation (Pilling et al. 2011).

Figure 8 presents the atom budget analysis performed as a function of fluence. The atomic column density evolutions for both precursors and products are compared. Data are calculated as follows:

(i) for each beam fluence selected for the FTIR monitoring, the column densities of hydrogen, carbon, oxygen and nitrogen atoms are determined for the most abundant daughter molecular species: CO, $CO_2$, $CH_4$, $C_2H_6$, $C_3H_6$, $CH_3OH$, $CH_3COOH$, $H_2CO$, $NH_3$, $N_2O_3$, $N_3$, NO, $NO_2$, HNCO, HCN and OCN. Molecular species like $C_2H_5NO_2$ were not taken into account because the assignments are not sure; the A-value of $O_2$ molecules dispersed in the ice is not known.

(ii) each molecular column density is multiplied by the number of hydrogen, carbon, oxygen or nitrogen atoms existing in the corresponding molecule, generating four ensembles of atomic column densities;

(iii) these column densities for each ensemble are summed up independently, yielding the minimum numbers of H, C, N or O atoms that should be delivered by precursors to form the observed products;

(iv) the sum obtained for each ensemble is divided by the corresponding number of atomic species existing in the precursor molecules - this represents the minimum column density of each precursor species necessary to form the products for that fluence. These four curves can be compared with $N_0 - N(F)$, the column density of the destroyed $CH_3COCH_3$ molecules (open stars) or with that of the destroyed $N_2$ molecules (crosses). Note that for these particular precursor ices, the atomic species are distinct in the two molecular species; this circumstance allows to compare directly the column densities of the products containing N with that of $N_2$ molecules.

## 5 DISCUSSION

Aiming to understand the radiolysis of acetone diluted in nitrogen, as well as the synthesis of daughter molecules, the infrared spectra of pure acetone irradiated by MeV ions at different fluences (Andrade et al. 2014) are compared with those of the (1:10) acetone:nitrogen mixture. Besides the obvious effect of lowering the acetone concentration when it is dissolved in nitrogen, distinct molecular dissociations are expected when each acetone molecule is surrounded by nitrogen molecules instead of other acetone molecules. Another important goal of the current experiment is to identify the new hybrid molecules formed from the two precursor species. This study is relevant from the point of the view of TNO ices, since acetone was recently discovered in a short-period comet, probably originating from the Kuiper Belt. It is also expected that acetone, a more complex molecule than $N_2$, occurs in a much lesser concentration than this one.

In the 90's, near-infrared analysis of Plutos and Tritons surfaces showed the existence of $CH_4$ diluted in $N_2$ matrix (Cruikshank et al. 1993; Owen et al. 1993). Other spectroscopic studies revealed that Pluto is also rich in CO. These three compounds are volatile at the surface temperatures of Pluto (30-50 K), being $N_2$ the most volatile species out of the three. $CH_4$, $N_2$, and CO can reach solid-gas phase equilibrium; moreover, the atmospheric gases may condensate on the icy surface and occult species contained in deeper layers, composed by others molecular species (Brown 2002; Cruikshank et al. 1997).

In order to clarify these processes, reactions involving species present in such environment need **to be** studied. The infrared absorption detection of the $N_2$ is not an easy task because it is a homo-nuclear diatomic molecule. Most of the time, the presence of $N_2$ in TNOs is deduced indirectly from small shifts in the wavelengths of some far infrared $CH_4$ absorption features (Tegler et al. 2008; Brown 2002). Moreover, the molecular atmosphere can also be lost. Volatiles will be preserved on objects only if these are cold or massive enough to prevent significant Jeans escape.

Only recently, in 2015, the NASA's New Horizons spacecraft visited Pluto and analyzed the colors and chemical compositions of their surfaces showing $H_2O$, $CH_4$, CO, $N_2$, and $NH_3$ ices as well as a reddish material which may be tholins (Cruikshank et al. 2015). At altitudes lower than 1800 km, molecular nitrogen ($N_2$) dominates, while methane ($CH_4$), acetylene ($C_2H_2$), ethylene ($C_2H_4$), and ethane ($C_2H_6$) are abundant minor species. These molecules likely produce an extensive haze that involves Pluto (Gladstoneet al. 2016; Bagenal et al. 2016).

### 5.1 Comparing acetone and acetone-nitrogen radiolysis

Comparing the current results with those for pure acetone (Andrade et al. 2014), it is observed that, under radiation, the $N_2$ column density decreases very fast; a possible interpretation is that nitrogen is quickly leaving a porous sample. Acetone presents a high compaction cross section ($1.6 \times 10^{-11}$ cm$^2$), suggesting that the ice formed by deposition of the vapor mixture is more porous than that of pure acetone ice.

Table 5. Column density variations of the parent molecules and of the observed products; destruction or formation molecular yields (number of molecules destroyed + sputtered or produced per projectile in $10^4$ for a fluence of $1.00 \times 10^{13}$ ions.cm$^{-2}$) atomic yields.

| Species | $\Delta N(F_{end})$ $\times 10^{17}$ | Molecular Yield $\times 10^4$ | Hydrogen Yield $\times 10^4$ | Carbon Yield $\times 10^4$ | Oxygen Yield $\times 10^4$ | Nitrogen Yield $\times 10^4$ |
|---|---|---|---|---|---|---|
| $CH_3COCH_3$ | −5.5 | 5.5 | 33.0 | 16.5 | 5.5 | – |
| $N_2$ | −329 | 329 | – | – | – | 658 |
| $\Sigma$ | – | – | 33.0 | 16.5 | 5.5 | 658 |
| CO | 1.30 | 1.30 | – | 1.30 | 1.30 | – |
| $CO_2$ | 0.42 | 0.42 | – | 0.42 | 0.84 | – |
| $CH_4$ | 0.93 | 0.93 | 3.72 | 0.93 | – | – |
| $C_2H_6$ | 1.40 | 1.40 | 8.4 | 2.8 | – | – |
| $C_3H_6$ | 2.95 | 2.95 | 17.7 | 8.85 | – | – |
| $CH_3OH$ | 0.13 | 0.13 | 0.52 | 0.13 | 0.13 | – |
| $CH_3COOH$ | 0.04 | 0.04 | 0.16 | 0.08 | 0.08 | – |
| $H_2CO$ | 0.19 | 0.19 | 0.38 | 0.19 | 0.19 | – |
| $N_3$ | 0.68 | 0.68 | – | – | – | 2.04 |
| $NH_3$ | 0.13 | 0.13 | 0.39 | – | – | 0.13 |
| NO | 0.003 | 0.003 | – | – | 0.003 | 0.003 |
| $NO_2$ | 0.02 | 0.02 | – | – | 0.04 | 0.02 |
| $N_2O_3$ | 0.06 | 0.06 | – | – | 0.18 | 0.12 |
| HNCO | 0.34 | 0.34 | 0.34 | 0.34 | 0.34 | 0.34 |
| HCN | 0.18 | 0.18 | 0.18 | 0.18 | – | 0.18 |
| OCN | 0.03 | 0.03 | – | 0.03 | 0.03 | 0.03 |
| $C_2H_5NO_2$ [a] | 0.002 | 0.002 | 0.01 | 0.004 | 0.004 | 0.002 |
| $\Sigma$ | – | – | 31.7 (96 per cent) | 13.9 (84 per cent) | 3.1 (56 per cent) | 2.9 (0.44 per cent) |

[a] Not considered in the budget calculation.

As far as destruction cross sections of acetone for both experiments are concerned, the values for pure and mixture ices are $3.6 \times 10^{-13}$ cm$^2$ and $2.5 \times 10^{-13}$ cm$^2$, respectively. It is concluded that, taking the same skeletal band (at 1229.0 cm$^{-1}$) as reference, acetone dissolved in nitrogen is 30% more difficult to be destroyed by the ion beam than the pure one is.

Dissolving acetone in nitrogen also affects the formation cross sections, $\sigma_f$, of daughter species: $CH_4$ and $H_2CO$, they are lower than the respective values for irradiated pure acetone, while for CO and $C_2H_6$ (ethane) species, the $\sigma_f$'s are higher. The most abundant species formed by radiolysis of pure acetone ice are: $CH_4$, CO, $H_2CO$, $CO_2$, $C_2H_4$, and $C_2H_6$, while in the nitrogen-acetone radiolysis, these molecules are also formed but the most abundant species are $C_3H_6$, $C_2H_6$, CO, $N_3$, $CO_2$ and $CH_4$. From the atom budget analysis, it is concluded that free hydrogen is rare in the ice. The $CH_2$ and $CH_3$ released in the mixture ice should then favor the $C_3H_6$ and $C_2H_6$ formation, rather than producing $CH_4$.

According to Bennett and colleagues (2006), in KBOs like Makemake which have the largest slabs of $CH_4$, the $C_2H_6$ molecule is generated from the union of two $CH_3$ radicals formed by the removal of a hydrogen atom from methane, being the first stable molecule to be formed during $CH_4$ irradiation. According to (Brown 2011), the $C_2H_6$ generation is blocked on Pluto due to the dilution of $CH_4$ in $N_2$ matrix. The removal of an H atom from $CH_4$ by radiolysis may occur, but the probability of existing another $CH_3$ radical around to generate $C_2H_6$ is low. The small amount of $C_2H_6$ found on Pluto (DeMeo, et al. 2009; Delsanti et al. 2010) originates possibly from the small regions of pure $CH_4$. The present results show that if acetone is present in Pluto, diluted in the $N_2$ matrix, $C_2H_6$ formation is facilitated.

### 5.2 Hybrid products of the acetone-nitrogen radiolysis

An overview of possible chemical reactions induced by the acetone-nitrogen radiolysis is presented in Fig. 9. The diagram is based on results obtained from acetone ice irradiated by Ni ions (Andrade et al. 2014) and by soft X-Ray photon induced desorption (Almeida et al. 2014), as well as from the current work. Almeida and colleagues investigated the photodesorption and the photo stability of pure acetone ices due to soft X-ray impact, presenting photodesorption yields for the formed positive ionic fragments at the O1s energy (~ 533 eV), and proposed that the Auger process is the main mechanism involved in the fragment desorption process. They found that the desorbed positive ions are: $CH_2^+$ (35.9%), $CO_2^+$ (18.7%), $O^+$ (9.3%), $CH_2CO^+$ (8.8%), $HCO^+$ and/or $C_2H_5^+$ (8.3%), $CH_3CO^+$ (6.7%), $C_2O^+$ (6.2%), $C_2CH_2^+$ (3.1%), $H^+$ (1.6%) and $CO^+$ (1.3%). These results indicate that the C-C bond is preferentially broken, forming $CH_2$. Oxygen is also abundantly delivered and should react with free CO, $CH_2$ and $CH_3$ species (however, this latter was not seen by Almeida and co-workers). The higher stability of the C-H bond provides a path to the $CH_2O$, $CH_2OH$ and $CH_3OH$ formation.

In the present work, the molecular nitrogen destruction cross section found in the ice mixture is 9.8 x the reason that the compaction cross section be relatively high. This fact can be a hint for the understanding of the nitrogen band absence in spectra of some comets where CHON molecules are present, like the 67P/Churyumov-

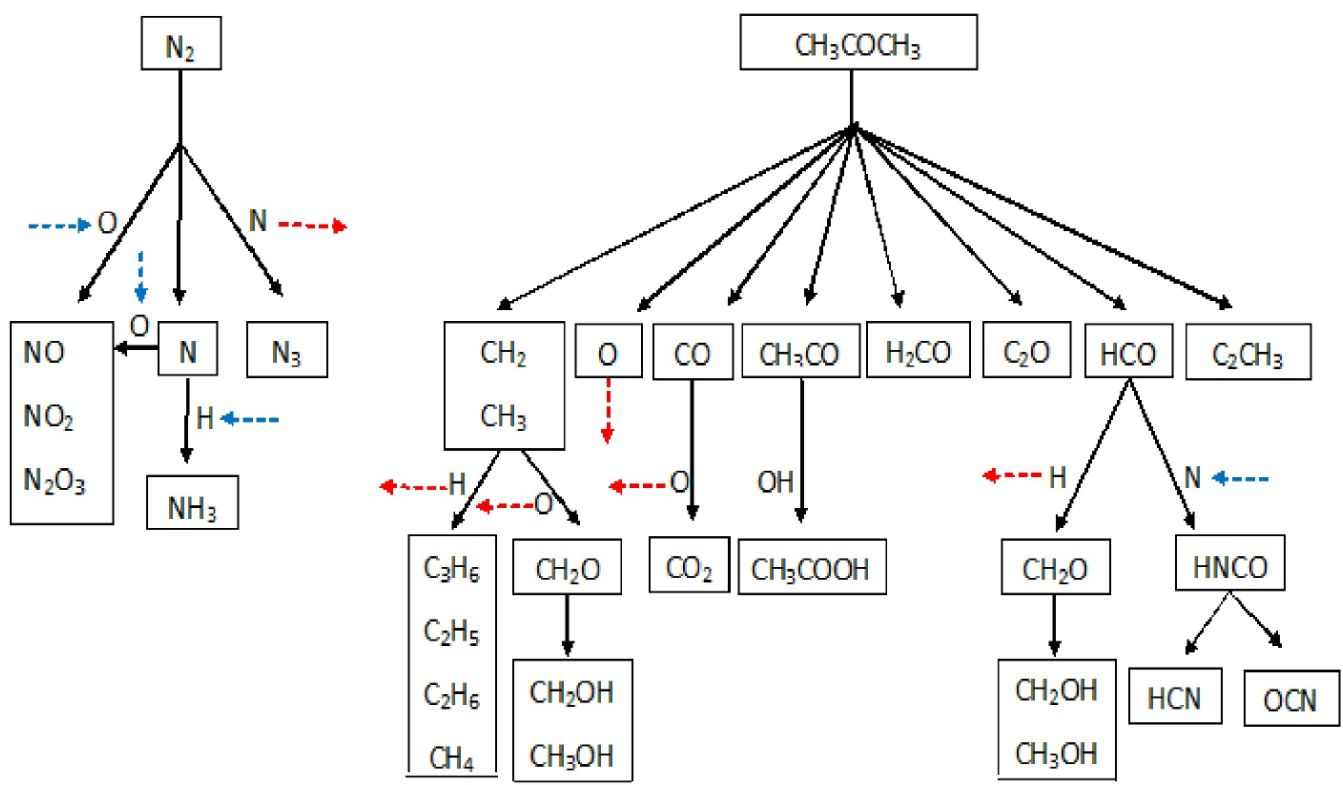

Figure 9. Possible pathways for daughter molecule formation due to $CH_3COCH_3:N_2$ radiolysis. Outgoing and ingoing dash arrows represent fragment atomic species delivered into and captured from the ice matrix, respectively.

$10^{-13}$ cm$^2$, corresponding to a destruction rate about three times higher than the acetone one. It is expected that a relatively large number of free N atoms be delivered in the matrix at the beginning of irradiation; indeed, at the highest projectile fluences, the $N_2$ IR peaks practically disappear whereas the acetone ones, remain in the spectra. The fast release of nitrogen is maybe

Nitrogen oxides are very stable and easily formed by radiolysis. For instance, FTIR analysis of the bombardment of $N_2$ + $H_2O$ ice mixture by heavy ions revealed that eight nitrogen oxides are formed (de Barros et al. 2015), among them $N_2O$ and $NO_2$ are the most abundant species. In two other systems, $H_2O:N_2:O_2$ and $CO:N_2$ mixtures, $NO_2$ (at 1616 cm$^{-1}$), NO (1874 cm$^{-1}$) and $N_2O$ (2236 cm$^{-1}$) were easily formed (Boduch et al. 2012; Sicilia et al. 2012). Thus, in astrophysical environments where C, N and O are present, these oxides are expect to be formed. However, the identification of their bands in the infrared spectra of space objects is not a simple task because the respective wavenumbers are too close to those corresponding to high abundant species such as water (at 1605 cm$^{-1}$) and CO (at 2234 cm$^{-1}$). Only the 1874 cm$^{-1}$ NO band is easily identified from infrared spectra of spatial sources because no strong band occurs close to it; because of this, out of these three species, only the NO species was found undoubtedly in comets. In this way, it is promising to search for this band in the infrared spectra of spatial ices, since chemical models predict that NO should be present on the interstellar grains (Charnley et al. 2001; Boduch et al. 2012). In this work, only $NO_2$ and NO were identified in the mixed ice, despite their very low abundance.

Gerasimenko (Altwegg et al. 2016). The current results also show that nitrogen has rather low reaction rate with acetone or with its daughters: very few hybrids species are formed (represented by dash lines in Fig 9). Even at low concentration, acetone seems to control the formation of new species.

The most abundant nitrogen-bearing species formed by the radiolysis of the analyzed mixture is $N_3$, followed by HNCO. The HCN, $NH_3$ and OCN species are also formed, but with lower efficiency.

New Horizons data showed that nitrogen-rich atmosphere of Pluto is escaping into space at a rate of about 1000 molecules per second, four orders of magnitude smaller than expected. This occurs, probably, because of the high cooling effect in its atmosphere. A possible coolant could be the hydrogen cyanide (HCN), which is very efficient and was recently detected in Pluto's atmosphere. The current results show that it is possible to form HCN and other N-bearing species from acetone diluted in $N_2$, suggesting that TNOs like Pluto and Triton - if containing acetone - could provide an endogenous source of HNC, HNCO, $NH_3$, OCN and glycine ($C_2H_5NO_2$).

### 5.3 Possible formation of glycine

In this work, the presence of glycine is investigated by several bands (Fig. 3(b) and (c)). However, the most intense bands of glycine are very close to some of acetone, turning difficult to definitely confirm the glycine formation via this pathway. Other amino acids were not found in the present work.

TNOs are supposed to be accreted from a mixture of rocks and volatile ices, which existed in the coldest regions of the proto-solar nebula. Today, most of them still keep their pristine form, preserving their composition and primordial structure. Hollis et al. (2015) suggest that short-lived radioactive isotopes like $^{26}$Al and $^{60}$Fe could played an important role in the evolution of Kuiper comets and their parents. During their nuclear decay, these radioisotopes provide an important source of heating, leading to the melting of water ice and triggering reactions; while long-lived isotopes should affect the largest objects. In this way and if the ingredients are present, formation of glycine is conceivable, although not proved.

Altwegg (2016) and co-workers reported surely, for the first time, the presence of volatile glycine in the coma of 67P/Churyumov-Gerasimenko. Recent studies pointed out how glycine could be formed in space from the species already existing there: Pilling et al. (2011) used soft X-ray photo-dissociation experimental data to calculate the enthalpy required to promote a possible scenario, in star-forming regions, for glycine formation via formic and acetic acids. They noted that, in gas phase, the destruction of formic acid by soft X-rays is faster than that of acetic acid, suggesting that the two most favorable reactions in the gas phase are $CH_3COOH + NH$ and $CH_3COOH + NH_2OH$. Other possible reactions include $NH_2CH_2$ and $COOH$, the carboxyl radical, one of the most abundant fragments produced by the photo-dissociation of acetic acid in their experiment. In solid phase, they suggested that, from the thermodynamical point of view, the reactions $HCOOH + NH_2CH$ and $HCOOH + NH_2CH_2OH$ are the two most favorable pathways to form glycine.

According to Chakrabarti et al. (2000), glycine can be easily formed in interstellar ices constituted by $H_2CO$, $HCN$ and $H_2O$. In the same way, but using quantum chemical calculations, Woon (2002) showed the viability of various pathways for the formation of glycine in the ice from HCN hydrogenation. The first molecule to be formed is $CH_2NH_2$, which in turn reacts with the $COOH$ produced by the $CO + OH$ reaction. Thus, the final pathway to form glycine in his set of reactions is $CH_2NH_2 + COOH \rightarrow C_2H_5NO_2$. Woon (2002) suggested that ice could have experienced thermal shocks or have been formed in comets while traversing warmer regions of the solar system. Under these conditions, glycine has a higher probability to be formed.

Besides of the discussion performed on HCN and its derivatives in the prebiotic evolution, some authors have shown that N-bearing molecules, in the presence of water and ammonium, can form DNA bases (such as adenine and guanine) and amino acids at temperatures as low as 77 K (Matthews 1995; Moore et al. 2002; Levy et al. 2000). The current results suggest that interesting prebiotic chemistry can occur from radiolysis of acetone diluted in a $N_2$ matrix, on the path of bio-molecule production.

## 6 REMARKS AND CONCLUSIONS

The radiolysis induced by 40 MeV nickel ions in the (10:1) $N_2$:$CH_3COCH_3$ ice mixture has been performed and analyzed. The destruction cross sections of both precursor molecules have been determined, as well as the formation and destruction cross sections of the daughter molecules.

Propane and ethane present the highest formation cross sections, followed by carbon monoxide, methane and carbon dioxide. The deposited ice mixture is quite porous, compacts rapidly during the irradiation and allows $N_2$ molecules to desorb quickly from the ice; nevertheless, daughter nitrogen containing species are observed. These species, such as HCN, $NH_3$ and, possibly glycine, are relevant for prebiotic chemistry.

An interesting finding about the presence of nitrogen around acetone molecules is the enhancement of $C_2H_6$ and the production of $C_3H_6$. An interpretation for this behavior is the formation of HNCO and $NH_3$, high consumers of free hydrogen in the ice matrix; the H-capture rate by $CH_3$ acetone fragments decreases and $CH_4$ production is inhibited. The abundant $CH_3$ radicals synthesize $C_2H_6$ and, after collision with free carbon, $C_3H_6$.

Results of the current experiment have implications on different astrophysical ices, mainly in solar system objects, since they mimic some effects of cosmic ray interaction with ices. At heliocentric distances larger than 30 AU, amorphous ices are formed and gases may be trapped in their pores. In the surfaces of the Trans-Neptunian Objects (TNOs) and of the Kuiper Belt comets, $N_2$ is probably present and mixed with other molecules, like water, carbon oxide, acetone and methanol. These surfaces are processed by cosmic rays, changing their chemical composition.

A question of particular interest is the formation of more complex molecules. In this direction, $C_2H_5NO_2$ and $CH_3COOH$ are synthesized in the system $CH_3COCH_3$-$N_2$ mixture ice exposed to energetic ions, being a possible stage in the complex process related to the origin of life.

## ACKNOWLEDGMENTS

This work was supported by the French-Brazilian exchange program CAPES-COFECUB. We are grateful to T. Been, C. Grygiel, A. Domaracka and J. M. Ramillon for their invaluable support. The Brazilian agencies CNPq (INEspaço), INCT-A, CNPq (304051/2014-4), CAPES (BEX 5383/15-3) and FAPERJ (E-26/110.087/2014, E-26/213.577/2015, E-26/216.730/2015 and E-05/2015-214814) are also acknowledged.